\documentclass[reprint,amsmath,amssymb,aps,superscriptaddress,hidelinks]{revtex4-2}
\usepackage[utf8]{inputenc}
\usepackage{graphicx}
\usepackage{hyperref}
\usepackage{dcolumn}
\usepackage{bm}
\usepackage{tcolorbox}
\usepackage{ulem}
\usepackage{xcolor}

\renewcommand{\sout}[1]{\bgroup\markoverwith{\textcolor{red}{\rule[0.5ex]{2pt}{1.5pt}}}\ULon{#1}}

\begin{document}

\title{Robustness of perfect transmission resonances to asymmetric perturbation}

\author{Ioannis Kiorpelidis}
\affiliation{LAUM, UMR-CNRS 6613, Le Mans Universit\'e, Avenue O. Messiaen, 72085, Le Mans, France}
\affiliation{Department of Physics, University of Athens, 15784, Athens, Greece}

\author{Panayotis Kalozoumis}
\affiliation{Department of Engineering and Informatics, Hellenic American University, 436 Amherst Street, Nashua, New Hampshire 03063, USA}

\author{Georgios Theocharis} 
\affiliation{LAUM, UMR-CNRS 6613, Le Mans Universit\'e, Avenue O. Messiaen, 72085, Le Mans, France}

\author{Vassos Achilleos}
\affiliation{LAUM, UMR-CNRS 6613, Le Mans Universit\'e, Avenue O. Messiaen, 72085, Le Mans, France}

\author{Fotios K. Diakonos}
\affiliation{Department of Physics, University of Athens, 15784, Athens, Greece}

\author{Vincent Pagneux} 
\affiliation{LAUM, UMR-CNRS 6613, Le Mans Universit\'e, Avenue O. Messiaen, 72085, Le Mans, France}

\begin{abstract}
We investigate the impact of asymmetric perturbations on the perfect transmission resonances (PTRs) of one-dimensional finite periodic systems. With no perturbations, the scattering region consists of $N$ identical cells, and the transmission spectrum exhibits at least $N-1$ PTRs in each pass band of the Bloch dispersion of the unit cell.  By introducing a perturbation, the periodic structure is broken, which \textit{a priori} results in the elimination of all PTRs.  However,  we demonstrate that PTRs can still arise under asymmetric perturbations when the unperturbed system possesses mirror symmetry, utilizing the $\mathcal{PT}$ symmetry of the unperturbed reflectionless eigenvalue problem. We also reveal an intriguing connection between two seemingly independent PTRs that lies in the symmetry of the unperturbed unit cell:  If one PTR is preserved, then a dual one is necessarily also preserved. Our findings offer insights for the design of, for example, a robust antireflection setup at multiple wavelengths or all-optical diode devices.
\end{abstract}

\maketitle


\section{Introduction}

Demonstrating and controlling optimal wave transmission properties in complex media are essential for various fields of study in wave physics with applications in photonic \cite{photonic,photonic2},  atomic \cite{atomic,atomic2}, phononic \cite{phononic}, and electronic \cite{electronic} systems, among others. Along these directions, perfect transmission resonances (PTRs) \cite{Zhukovsky2010}, which correspond to particular frequencies where the modulus of the transmission coefficient is 1, play a key role in the transmission spectrum of a scattering setup \cite{Soukoulis2010}.  PTRs are usually supported in  systems that possess some kind of symmetry, for instance, global mirror  \cite{Gong1999,Hu2001,Mazzer2002,Mauriz2009} or local \cite{Kalozoumis2013a, Kalozoumis2013b} symmetry. In this framework, one-dimensional finite periodic setups have been intensively studied \cite{Martorell1993}, and when a cell is repeated $N$ times in space, then the transmission spectrum of the setup shows bands with at least $N-1$ PTRs \cite{Martorell1993,Griffiths1992,Griffiths2001,Barra1999}. Additionally, an interesting property  is that if the cell possesses mirror symmetry, as in the case, for instance, of finite Kronig-Penney models, then global mirror-symmetry conditions are imposed on the scattering states corresponding to PTRs \cite{Pereyra2017}.

As soon as a scattering system is perturbed, its PTRs are typically lost due to the breaking of symmetry. However, it has been shown that PTRs can also arise in asymmetric systems \cite{Zhukovsky2010, Naumis2009, Lu2005}, highlighting that symmetry is not a necessary condition. Along this line, achieving high transmission in disordered media has attracted significant attention over the years. Notable developments, especially in photonics, include the emergence of necklace states driven by resonant effects \cite{Wiersma2005, Toninelli2015}, the creation of constant-intensity waves by tuning gain and loss in non-Hermitian systems \cite{Makris2015, Park2018}, and the inverse design of eigenstates by precisely controlling specific parameters \cite{Park2016}. Here, we focus on unrevealing how to preserve PTRs by asymmetric perturbations in prototypical one-dimensional (1D) setups, where it is possible to carry out explicit asymptotic analysis, which in turn provides constructive design of the perturbations.

In particular, in our work, we explore the impact of asymmetric perturbations on the transmission spectrum of a 1D finite periodic scattering system that is built from a mirror symmetric cell. By calculating the influence of the perturbation on the frequencies that correspond to PTRs, we determine how the design of particular asymmetric perturbations, made of a series of Dirac delta scatterers, can protect a desired PTR. Additionally, owing to the mirror symmetry of the unit cell, we prove that if the perturbing Dirac scatterers protecting a PTR (say, a number $n$ between 1 and $N-1$) are placed either at the centers or at the edges of the cell, then a dual PTR (number $N-n$) is protected as well. 

Our work is organized as follows: In Sec. II, we provide a brief review of scattering by a finite periodic system that is built from a mirror-symmetric cell, and we recall the appearance of PTRs in the transmission spectrum. In Sec. III, we consider a perturbation in the scattering region, and we calculate the first-order correction to the frequencies that correspond to the PTRs. Then, we show how we can design a perturbation that consists of Dirac scatterers and preserves a desired PTR. In Sec. IV, we place Dirac scatterers either at the centers or at the edges of the cells, and we prove that if a PTR is preserved, then a dual PTR is preserved as well. In Sec. V, we discuss the asymmetry of the scattering states corresponding to perturbed PTR, and in Sec. VI we explore the impact of the perturbation strength to PTRs. Finally, in Sec. VII we summarize our findings.


\section{Review of scattering by a periodic system with mirror-symmetric cells}
\label{Section2}
We begin by recalling some basic properties of the wave scattering in one dimension by a setup that is finite periodic and mirror symmetric. We consider waves satisfying the stationary Schr\"{o}dinger equation
\begin{equation}
\psi'' + \left[ k^2 - \mathcal{V}_0(x) \right] \psi =0 ,
\label{eqShrodinger}
\end{equation}
where the prime denotes differentiation with respect to space coordinate $x$, $\psi$ is the wave function, $k$ is the frequency, and $\mathcal{V}_0(x)$ is the potential, which is nonzero only for $x \in [-D/2,D/2]$.

\begin{figure}
\begin{center}
\includegraphics[width=1\columnwidth]{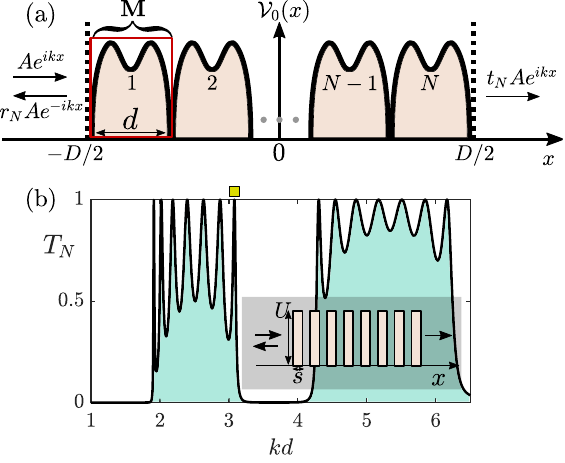}
\end{center}
\caption{ 
(a) Schematic description of the scattering of a wave with amplitude $A$ and frequency $k$ by a finite periodic setup that consists of $N$ mirror-symmetric cells. 
The length of each cell is $d$, and that of the whole setup is $D$, where $D=Nd$.
Also, $t_N$ ($r_n$) is the transmission (reflection) amplitude.
(b) Transmission spectrum $T_N(k)=|t_N(k)|^2$ of a finite periodic scattering setup that consists of eight rectangular barriers (the scattering setup is shown in the inset).
The length of each barrier is $s=d/6$, and therefore, the free space between two neighboring barriers has length $5d/6$ (note that the setup in the inset does not represent such distances). 
The heights of the barriers, $U=\displaystyle \max_x V_0(x)$, are set at $\sqrt{U} \approx 5.196/d$.
}
\label{fig1}
\end{figure}

The potential $\mathcal{V}_0(x)$ is assumed to be real and finite periodic as described in Fig.~\ref{fig1}(a).
For reasons that will be clear later, we choose a mirror-symmetric unit cell.
Such a potential $\mathcal{V}_0$ is written in the following form:
\begin{equation}
\mathcal{V}_0(x)
=
\sum_{n=1}^N {V}_0 \left( x+\frac{d}{2}(N+1)-nd \right),
\label{eqpotential}
\end{equation}
where $V_0(x)=V_0(-x)$ is real and zero outside the region $[-d/2,d/2]$.

The transfer matrix of the building cell takes the form \cite{Barriuso2012}
\begin{equation}
\mathbf{M}
=
\begin{pmatrix}
1/t^* && r/t
\\
-r/t && 1/t
\end{pmatrix}
,
\label{eqtransfer_one_cell}
\end{equation}
where $t$ ($r$) is the transmission (reflection) amplitude and, since $V_0$ is real, $T+R=1$, where $T=|t|^2$ ($R=|r|^2$) is the transmission (reflection) coefficient of the cell.
The transfer matrix of the periodic setup with $N$ cells, $\mathbf{M}^N$, is given by 
\begin{equation}
\mathbf{M}^N
=
\begin{pmatrix}
1/t_N^* && r_N/t_N
\\
-r_N/t_N && 1/t_N
\end{pmatrix},
\end{equation}
where $t_N$ ($r_N$) is the transmission (reflection) amplitude of the $N$ cells.
Similar to the single-cell case, $T_N+R_N=1$ where $T_N=|t_N|^2$ and $R_N=|r_N|^2$ are now the transmission and reflection coefficients of the $N$ cells.
Owing to the Chebyshev identity \cite{Soukoulis2010}, one can write $\mathbf{M}^N$ as
\begin{equation}
\mathbf{M}^N
=
\tiny
\begin{pmatrix}
\dfrac{1}{t^*} \dfrac{\sin(N \phi)}{\sin(\phi)} - \dfrac{\sin \left[ (N-1) \phi \right] }{\sin(\phi)}  
&& 
\dfrac{r}{t} \dfrac{\sin(N \phi)}{\sin(\phi)}
\\
-\dfrac{r}{t} \dfrac{\sin(N \phi)}{\sin(\phi)}
 && 
 \dfrac{1}{t} \dfrac{\sin(N \phi)}{\sin(\phi)} - \dfrac{\sin \left[ (N-1) \phi \right] }{\sin(\phi)}  
\end{pmatrix}
\normalsize,
\label{eqtransfer_N_cells}
\end{equation}
where $\phi=\dfrac{1}{2} \text{Tr}[\mathbf{M}]$ is the Bloch phase of the unit cell.

Since we are focusing on PTRs in finite periodic setups, a useful expression for the transmission coefficient of the $N$ cells is \cite{Martorell1993,Griffiths1992,Griffiths2001,Soukoulis2010}
\begin{equation}
T_N=\displaystyle{
\dfrac{1}{1+\left( \dfrac{1}{|t|^2}-1 \right) \dfrac{\sin^2(N\phi)}{\sin^2(\phi)}}}
.
\label{eq5}
\end{equation}

In Eq.~\eqref{eq5} PTRs, i.e., $T_N=1$, are obtained when $\phi_n=n\pi/N+\mod(2\pi)$, with $n=1,2,...,N-1$, since $\sin(N\phi_n)=0$. Additional PTRs appear as well whenever $T=1$.
Figure~\ref{fig1}(b) displays the transmission spectrum $T_N(k)$ of a periodic setup with eight rectangular barriers (the setup is shown in the inset). 
The transmission spectrum shows a bandlike structure with $N-1=7$ PTRs in each band since there is no $k$ for which  $T(k)=1$ in this frequency range.

As a remark, we note that PTRs correspond to real eigenvalues of the reflectionless mode eigenproblem \cite{Pagneux2018, Stone2020}, given by the solution of Eq.~\eqref{eqShrodinger} with boundary conditions
\begin{equation}
\left. \dfrac{d\psi(x)}{dx}\right|_{x=\pm D/2} - ik\psi(\pm D/2) = 0
,
\label{eqboundary1}
\end{equation}
which are known as Robin boundary conditions \cite{Siegl2011}.
One property of this reflectionless eigenvalue problem defined by Eqs.~\eqref{eqShrodinger} and \eqref{eqboundary1} is that it becomes $\mathcal{PT}$ symmetric when the potential $\mathcal{V}_0(x)$ is mirror symmetric $\mathcal{V}_0(x)=\mathcal{V}_0(-x)$ (see Appendix \ref{appendixA} for more details).
The latter suggests that perturbing the symmetric $\mathcal{V}_0(x)$ and keeping the mirror symmetry, PTRs are protected (but are displaced in the transmission spectrum) as long as they do not coalesce at the exceptional points of the $\mathcal{PT}$-symmetric reflectionless eigenvalue problem.
In contrast, when the mirror symmetry of the potential is broken, no PTR protection is anticipated, and we will inspect how non-mirror-symmetric perturbations are capable of asymptotically keeping desired PTRs.


\section{Perturbing the potential}
\label{Section3}

We consider that $\mathcal{V}_0(x)$ is perturbed and the scattering region is described by the potential 
\begin{equation}
\mathcal{V}(x)=\mathcal{V}_0(x) + \epsilon \mathcal{V}_1(x),
\label{perturbed_potential}
\end{equation}
where $\mathcal{V}_1(x)$ is the perturbation and $\epsilon$ is a small parameter, i.e., $\epsilon \ll 1$.
As for $\mathcal{V}_0(x)$, we let the perturbing potential $\mathcal{V}_1(x)$ be nonzero only inside the region $[-D/2,D/2]$.

\begin{figure}[t!]
\begin{center}
\includegraphics[width=0.85\columnwidth]{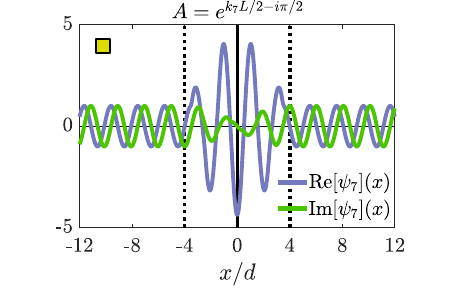}
\end{center}
\caption{ 
Real and imaginary parts of the wave function $\psi_7(x)$ for frequency $k_7$ (corresponding to PTR number $n=7$) of the unperturbed setup that is shown in Fig.~\ref{fig1}(b) (indicated with a yellow square). 
The amplitude of the incoming wave is set at $A=e^{ik_7 H/2 -\pi/2}$ [see Eq.~\eqref{eqsymmetric_phase}].
}
\label{fig2n}
\end{figure}

\subsection{First-order correction}
We aim to find the variation at the frequencies that correspond to PTRs.
We denote the unperturbed frequencies corresponding to $\mathcal{V}_0$ as $k_{0,n}$ (where the index $n$ is the number of a PTR in one passband, $n=1,2,...,N-1$).
Following the asymptotic perturbation approach of reflectionless modes in ref.~\cite{Pagneux2024},
we write the perturbed frequencies as 
\begin{equation}
k_n=k_{0,n}+\epsilon k_{1,n} + \ldots ,
\label{eq7}
\end{equation}
and by applying the classical solvability condition we find that the first-order correction $k_{1,n}$ is given by (see Appendix~\ref{appendixB} for the details)
\begin{small}
\begin{equation}
k_{1,n}
=
\dfrac{\int_{-D/2}^{D/2} \mathcal{V}_1 \psi^2_{0,n} dx}{i[ \psi_{0,n}^2(D/2)-\psi_{0,n}^2(-D/2)]+2k_{0,n} \int_{-D/2}^{D/2} \psi^2_{0,n} dx}.
\label{eqwave_number_corr}
\end{equation}
\end{small}In the last expression $\psi_{0,n}$ is the PTR wave function for frequency $k_{0,n}$.
To asymptotically keep a PTR at order $O(\epsilon)$ it is sufficient to look for a zero imaginary part of $k_{1,n}$.
Then, the real part of $k_{1,n}$ shows how this PTR is shifted in the transmission spectrum.

From Eq.~\eqref{eqwave_number_corr} it appears that we just have to deal with overlap integrals involving $\mathcal{V}_1$ and $\psi_{0,n}$, where the symmetry of $\psi_{0,n}$ plays an important role.
In fact, for a mirror-symmetric potential $\mathcal{V}_0$ like the one illustrated in Fig.~\ref{fig1}(a), great simplification can be obtained.
Indeed, with the suitable phase transformation described below, the real and imaginary parts of $\psi_{0,n}$ become symmetric and antisymmetric around $x=0$, respectively (see, for example, Fig.~\ref{fig2n}).
In that case, the denominator in Eq.~\eqref{eqwave_number_corr} is real, and therefore, the imaginary part of $k_{1,n}$ is given by
\begin{equation}
\text{Im}[k_{1,n}] =  \mathcal{C} \int_{-D/2}^{D/2} \mathcal{V}_1 \text{Re}[\psi_{0,n}] \text{Im}[\psi_{0,n}] dx ,
\label{eqwave_numer_imaginary}
\end{equation}
with $\mathcal{C}$ being real.
Considering scattering from the left side of the potential $\mathcal{V}_0(x)$ [see Fig.~\ref{fig1}(a)], the phase that symmetrizes the wave function is obtained by setting the amplitude $A$ of the incoming wave equal to
\begin{equation}
A=
\begin{cases}
e^{ik_{0,n} D/2},~~~~~~~~~\text{for even}~n,
\\
e^{ik_{0,n} D/2-i\pi/2}, ~~ \text{for odd}~n.
\end{cases}
\label{eqsymmetric_phase}
\end{equation}

\begin{figure}
\begin{center}
\includegraphics[width=1\columnwidth]{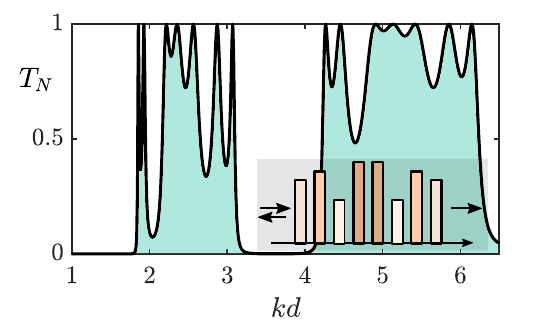}
\end{center}
\caption{
Mirror-symmetric perturbation ($\epsilon=0.1$).
Shown is the transmission spectrum of the scattering setup that is displayed in the inset.
This scattering setup is derived by perturbing in a mirror-symmetric way the barriers of the setup shown in Fig.~\ref{fig1}(b).
The lengths of the barriers and the distances between two neighboring barriers are the same as in Fig.~\ref{fig1}(b).
The perturbations $U_i$ in each barrier $i=1,2,...,8$ are set at $U_1=U_8=-27/d^2$, $U_2=U_7 = 0$,  $U_3=U_6 = -72/d^2$, and ${U}_4 ={U}_5 =27/d^2$.
}
\label{fig2}
\end{figure}

As mentioned before, in mirror-symmetric perturbations the PTRs are kept.
That means that if the perturbing potential $\mathcal{V}_1$ is an even function with respect to $x=0$ (mirror-symmetric perturbation), then from Eq.~\eqref{eqwave_numer_imaginary} we get $\text{Im}[k_{1,n}]=0$ for all $n$ since $\text{Re}[\psi_{0,n}] \text{Im}[\psi_{0,n}]$ is odd for all $n$. 
In Fig.~\ref{fig2} we present such an example.
In the following we show how we can design an asymmetric perturbing potential $\mathcal{V}_1$ that preserves PTRs.

\subsection{Preserving one PTR with asymmetric perturbation}
\begin{figure}
\begin{center}
\includegraphics[width=1\columnwidth]{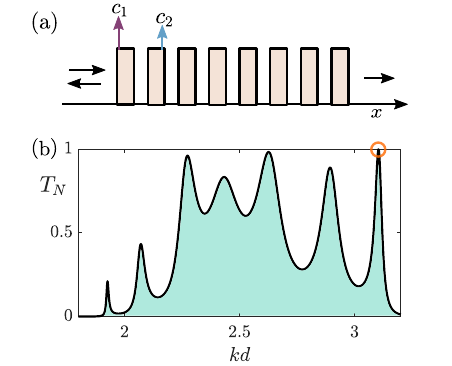}
\end{center}
\caption
{
Asymmetric perturbation preserving one PTR ($\epsilon=0.1$).
(a) 
Schematic description of the wave scattering by a setup that consists of eight rectangular barriers (unperturbed potential $\mathcal{V}_0$) and two Dirac scatterers (perturbing potential $\mathcal{V}_1$). 
The heights of the barriers $U$, their lengths $s$, and the free space between two neighboring barriers are the same as in Fig.~\ref{fig1}(b).
The first (second) Dirac scatterer is placed at a distance of $0.1 s$ ($0.95 s$) from the left edge of the first (second) barrier.
The strength of the first Dirac scatterer is set at $c_1=4.8/d$, and the strength of the second Dirac scatterer is found from Eq.~\eqref{eq_wave_number__Dirac_preserved}, so that the PTR number $n=7$ of the unperturbed setup is preserved after the addition of the two scatterers.
(b)
Transmission spectrum of the setup that is shown in (a).
Only the PTR number $n=7$ of the unperturbed setup  is preserved (indicated with a circle).
}
\label{fig4n}
\end{figure}

Choosing $\mathcal{V}_1$ to be a sum of Dirac scatterers
\begin{equation}
\mathcal{V}_1 = \sum_{m=1}^M c_m \delta(x-x_m),
\label{eqperturbing_Dirac}
\end{equation}
where $c_m$  are the strengths of the scatterers and $x_m \in [-D/2,D/2]$ are their positions, Eq.~\eqref{eqwave_numer_imaginary} takes the simple form
\begin{equation}
\text{Im}[k_{1,n}] 
=
\mathcal{C}
 \sum_{m=1}^M c_m \text{Re}[\psi_{0,n}(x_m)] \text{Im}[\psi_{0,n}(x_m)]
.
\label{eqwave_number_Dirac}
\end{equation}
Based on this form, we can achieve $\text{Im}[k_{1,n}]=0$ for a specific PTR number $n$ with asymmetric $\mathcal{V}_1$.
Actually, since $\psi_{0,n}(x)$ is given,  $\text{Im}[k_{1,n}]=0$ takes the form of one equation with $M$ unknowns ($c_m$, $m=1,...,M$),
\begin{equation}
\sum_{m=1}^M c_m \text{Re}[\psi_{0,n}(x_m)] \text{Im}[\psi_{0,n}(x_m)]
 =0
.
\label{eq_wave_number__Dirac_preserved}
\end{equation}
To comply with this equation, the minimal number of perturbing Dirac scatterers is generically $M=2$, resulting in a simple proportionality between $c_1$ and $c_2$ (accidental location where $\text{Re}[\psi_{0}] \text{Im}[\psi_{0}]=0$ would allow $M=1$).
As an illustration, consider the scattering by the potential shown in Fig.~\ref{fig4n}(a).
With $N=8$, it displays how to keep PTR $n=7$ with two Dirac scatterers.
Note that only the PTR $n=7$ is kept, while all the 6 other are lost.

We remark that we might generalize this procedure by increasing the number of perturbing Dirac scatterers,
allowing us to preserve a larger number of PTRs.
In principle, to keep $q$ PTRs we need $q+1$ Dirac scatterers.
In the next section, we show that due to the mirror symmetry of the unit cell, two PTRs can be preserved with only two Dirac perturbations when properly located.

\section{Exploiting the mirror symmetry to keep PTRs in pairs}
\label{section4}
When the unperturbed unit cell is mirror symmetric, particular properties come out for perturbing Dirac scatterers, all located either  at the centers or at the edges of the cells of $\mathcal{V}_0$.

\subsection{Dirac scatterers at the centers of the cells}
\begin{figure}
\begin{center}
\includegraphics[width=1\columnwidth]{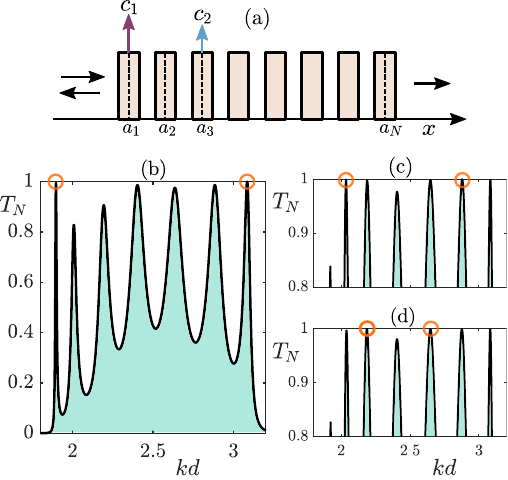}
\end{center}
\caption{ 
PTRs preserved in pairs: Dirac scatterers located at the centers of the cells ($\epsilon=0.1$).
(a) 
Schematic description of the wave scattering by a setup that consists of eight rectangular barriers ($\mathcal{V}_0$) and 2 Dirac scatterers  ($\mathcal{V}_1$) of strengths $c_{1,2}$ that are placed at the centers of the first and third cells.
Again, the lengths of the barriers $s$, the free space between two neighboring barriers, and their heights $U$ are the same as in Fig.~\ref{fig1}(b).
(b)-(d) Transmission spectrum of the setup that is shown in (a). 
(b) Specifically, the strength $c_1$ is set at  $c_1=12/d$.
The strength $c_2$ is designed through Eq.~\eqref{eq_wave_number__Dirac_preserved} so that the PTR number $n=1$ of $\mathcal{V}_0$ is preserved after the addition of the two scatterers.
The PTR number $n=7$ is also preserved.
(c) We set $c_1=4.5/d$, and we calculate the strength $c_2$ so that the PTR number $n=2$ of $\mathcal{V}_0$ is preserved. 
The PTR number $n=6$ is preserved as well.
(d) We set $c_1=2.4/d$, and we design $c_2$ so that the PTR number $n=3$ of $\mathcal{V}_0$ is preserved. 
Then the PTR number $n=5$ is also preserved.
}
\label{fig4}
\end{figure}

Let us start with the presentation of an example:
An incident wave is scattered by the potential shown in Fig.~\ref{fig4}(a).
This potential consists of $N=8$ rectangular barriers ($\mathcal{V}_0$) and $M=2$ Dirac scatterers ($\mathcal{V}_1$) that are placed at the centers of the first and third barriers.
In Figs.~\ref{fig4}(b)-(d) we illustrate the transmission spectrum initially designed to protect one PTR (either PTR number $n=1$, $n=2$ or $n=3$).

In Fig.~\ref{fig4}(b) we design the strengths $c_1$ and $c_2$ [from  Eq.~\eqref{eq_wave_number__Dirac_preserved}] so that the PTR  $n=1$ is preserved.
Surprisingly, it appears that in addition to the PTR $n=1$, the PTR $n=7$ is preserved as well.
In Figs.~\ref{fig4}(c) and \ref{fig4}(d) the same kind of behavior is observed: In designing the perturbation protecting PTR number $n$, the PTR number $N-n$ is also protected.

In fact, it can be proven that
if a periodic potential with mirror-symmetric cells (not necessarily rectangular cells) is perturbed by Dirac scatterers placed at the centers of the cells, then the PTRs are preserved in pairs.
More specifically, we show that 
\begin{equation}
\text{Im}[k_{1,n}]=0
\Rightarrow
\text{Im}[k_{1,N-n}]=0
\label{eqproperty}
\end{equation} 
for perturbations at the centers of the cells.
A sketch of the proof is as follows:
\begin{itemize}
\item In Appendix \ref{AppendixC} we prove that for the centers $a_p$ of the cells ($p=1,2,...,N$) 
\begin{equation}
\text{Re}[\psi_{0,n}(a_p)] \text{Im}[\psi_{0,n}(a_p)]
=f(\phi_n) g(p,\phi_n)
\label{eqReImcenters}
\end{equation}
where
\begin{equation}
g(p,\phi_n) = \sin^2\left[ (N-p)\phi_n \right] - \sin^2\left[ (p-1)\phi_n \right] 
\label{eqgcenters}
\end{equation}
and 
\begin{equation}
\phi_n = \dfrac{n \pi }{N}
\label{eqBloch}
\end{equation}
is the Bloch phase of the $n^{th}$ PTR.
\\
\item Equations \eqref{eqgcenters} and \eqref{eqBloch} suggest that under the  interchange $n \rightarrow N-n$,
\begin{equation}
\phi_{N-n}=\pi -\phi_n 
\Rightarrow
g(p,\phi_n) = g(p,\phi_{N-n}).
\label{eqggcenters}
\end{equation}
Using Eqs.~\eqref{eqReImcenters} and \eqref{eqggcenters}, we can show that if Eq.~\eqref{eq_wave_number__Dirac_preserved} holds for $n$, it holds for $N-n$ as well, implying property \eqref{eqproperty}.
\end{itemize}
More details can be found in Appendix \ref{AppendixC}.


\subsection{Dirac scatterers at the edges of the cells}
In the previous section we considered the case of Dirac scatterers located at the centers of cells.
Here, the Dirac scatterers are placed at the edges of cells.
The results in the previous section still apply, while an additional property is revealed.

\begin{figure}
\begin{center}
\includegraphics[width=1\columnwidth]{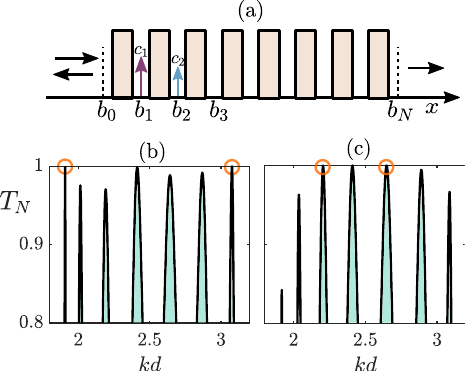}
\end{center}
\caption{
Protection of PTRs when Dirac scatterers are placed at the edges of the cells ($\epsilon=0.1$).
(a) 
Schematic description of the scattering  of a wave by a setup that consists of eight rectangular barriers ($\mathcal{V}_0$) and two Dirac scatterers  ($\mathcal{V}_1$) of strengths $c_{1,2}$ that are placed at the right edges of the first and second cells.
The heights $U$ of the barriers, their lengths $s$, and the free space between two neighboring barriers are as in Fig.~\ref{fig1}(b).
(b) and (c) Transmission spectrum of the setup that is shown in (a):
(b) We set the strength $c_1$ of the first Dirac scatterer equal to $c_1=1.8/d$,  and we calculate $c_2$ from Eq.~\eqref{eq_wave_number__Dirac_preserved} so that the PTR $n=1$ is preserved.
We find that the PTR number $n=7$ is also preserved. 
(c) We set $c_1=1.8/d$, and we calculate $c_3$ again from Eq.~\eqref{eq_wave_number__Dirac_preserved}, so that the PTR number $n=3$ is preserved.
We find that the PTR number $n=5$ is preserved as well.
}
\label{fig6}
\end{figure}

In line with our previous approach, we begin by illustrating an example.
We consider the wave scattering by the potential shown in Fig.~\ref{fig6}(a).
As before, the unperturbed potential $\mathcal{V}_0$ consists of $N=8$ rectangular barriers.
We place $M=2$ Dirac scatterers of strengths $c_{1,2}$ at the right edges of the first and second cells.
Unsurprisingly, by choosing $c_{1,2}$ according to Eq.~\eqref{eq_wave_number__Dirac_preserved} the PTRs number $n$ and $N-n$ are simultaneously preserved,
as shown in the transmission spectra in Figs.~\ref{fig6}(b) and \ref{fig6}(c) \cite{comment1}.
This property (also true for the setup in Sec. \ref{section4} A) appears for any even or odd $N$.

\begin{figure}
\begin{center}
\includegraphics[width=1\columnwidth]{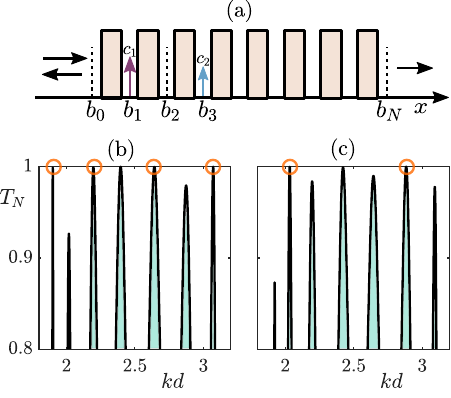}
\end{center}
\caption{
Same as in Fig.~\ref{fig6}, but now the two Dirac scatterers are placed at edges $b_1$ and $b_3$.
(b) and (c) We set $c_1=1.5/d$, and we calculate $c_2$ from Eq.~\eqref{eq_wave_number__Dirac_preserved}.
The PTRs in (b) are protected in multiplets.
}
\label{fig5}
\end{figure}

What is surprising is the following: If the number of cells $N$ is even and the Dirac scatterers are placed at the $b_p$'s with the same parity
--- $p$ can take only the values $p=1,3,5,...$ (or $p=0,2,4,...$) ---  so that the  PTR number $n$ is protected, 
then  not only the PTR number $N-n$ but also the PTRs number $N\pm n/2$ gets protected.
To illustrate this, we consider the scattering using the setup shown in Fig.~\ref{fig5}(a), where the two Dirac scatterers are placed at edges $b_1$ and $b_3$.
Figures~\ref{fig5}(b) and \ref{fig5}(c) show the transmission spectra for two choices $c_{1,2}$
where the perturbation is designed to protect the PTR $n=1$ and $n=2$, respectively.
Note that in Fig.~\ref{fig5}(b), by protecting one PTR, we actually protect three more according to the previous discussion.

We now show analytically all the aforementioned connections between PTRs.
That is, we show that when the Dirac scatterers are placed at the edges $b_p$  of cells, property \eqref{eqproperty} still holds, namely, $\text{Im}[k_{1,n}]=0
\Rightarrow
\text{Im}[k_{1,N-n}]=0
\label{eqproperty1}$
.
Moreover, if the scatterers are placed at the $b_p$’s with the same parity, i.e., $p=1,3,5$ (or $p=0,2,4,...$), 
the additional property
\begin{equation}
\text{Im}[k_{1,n}]=0
\Rightarrow
\text{Im}[k_{1,N/2 \pm n}]=0
\label{eqproperty2}
\end{equation}
also holds.
As we did in the previous section, we outline the steps of the proof; details can be found in Appendix \ref{AppendixD}:
\begin{itemize}
\item In Appendix \ref{AppendixD} we show that at the edges $b_p$ of the cells ($p=0,1,2,...,N$) the following relation holds:
\begin{equation}
\text{Re}[\psi_{0,n}(b_p)] \text{Im}[\psi_{0,n}(b_p)]
=\tilde{f}(\phi_n) \tilde{g}(p,\phi_n),
\label{eqReImedges}
\end{equation}
where
\begin{equation}
\tilde{g}(p,\phi_n)=\sin(2p\phi_n)
\label{eqgedges}
\end{equation}
and $\phi_n$ is the Bloch phase of the $n$th PTR [see Eq.~\eqref{eqBloch}].
\\
\item Under the interchange $n \rightarrow N-n$, from  Eq.~\eqref{eqgedges} we find that 
\begin{equation}
\tilde{g}(p,\phi_n)=\tilde{g}(p,\phi_{N-n}).
\label{eqggedges}
\end{equation}
We can prove Eq.~\eqref{eqproperty} using Eq.s~\eqref{eq_wave_number__Dirac_preserved}, \eqref{eqReImedges}, and \eqref{eqggedges} as we did in the previous section.
\\ 
\item Under the interchange $n \rightarrow N/2 \pm n$, from  Eq.~\eqref{eqgedges} we obtain
\begin{equation}
\tilde{g}(p,\phi_{N/2 \pm n})=(-1)^p \tilde{g}(p,\phi_n).
\label{eqggedges2}
\end{equation}
Note that if $N$ is even and $p$ is only odd or only even, then we can prove property \eqref{eqproperty2} using Eqs.~\eqref{eq_wave_number__Dirac_preserved}, \eqref{eqReImedges}, and \eqref{eqggedges2}.
\end{itemize}


\section{Asymmetry of the wave field}
Up to this point, we have studied the variation of the frequencies $k_{0,n}$ under perturbations [recall Eq.~\eqref{eq7}].
We now focus on the perturbed wave field
\begin{equation}
\psi_n=\psi_{0,n}+\epsilon \psi_{1,n}+\ldots ~.
\label{perturbed_field}
\end{equation}
We will explore whether an asymmetric perturbation protecting the PTR number $n$
breaks the mirror symmetry of $|\psi_{0,n}|$, thus inducing an asymmetric wave field $\psi_n$.
This is interesting because it enables the appearance of nonreciprocal scattering when appropriate nonlinearity is added, 
leading to an asymmetric transmission  \cite{{Rodriguez2024}}  and even to all-optical diodes \cite{Zhukovsky2010, Zhukovsky2011, Biancalana2010, Biancalana2008}.  

\begin{figure}
\begin{center}
\includegraphics[width=0.9\columnwidth]{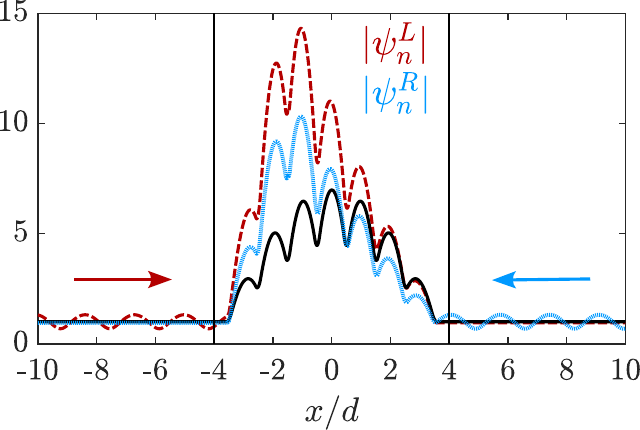}
\end{center}
\caption
{
The scattering setup is the one shown in Fig.~\ref{fig4}(a) for $\epsilon=0.225$.
The red dashed (blue dotted) line denotes the wave-field norm
for frequency $k_1 \approx 1.86696 / d$ (corresponding to the first peak of the frequency spectrum)
when scattering from the left (right) side of the setup is considered.
The black solid line represents $|\psi_{0,n}|$ for $n=1$.
}
\label{fig7}
\end{figure}

To this end, we use the asymmetric setup shown in Fig.~\ref{fig4}(a) with Dirac scatterers designed to protect the PTR $n=1$, and
we consider  scattering either from the left  or from the right side of this perturbed setup.
In Fig.~\ref{fig7} we show the corresponding norms of the exact fields $\psi_n^L$ and $\psi_n^R$ (in the presence of the perturbation for $n=1$) with dashed and dotted lines, respectively. With the arrows we indicate the direction of the incident waves. For comparison, we also show with a solid line the unperturbed $|\psi_{0,n}|$ for $n=1$.

Note that the fields $\psi_n^L$ and $\psi_n^R$ are asymmetric and close to resonance (transmission coefficient $T_N \approx 0.9$).
Moreover, we note that the asymmetry of $\psi_n$ grows as the value of the perturbation parameter $\epsilon$ increases (a large value of $\epsilon$ suggests the perturbed setup is strongly asymmetric).
However, for large $\epsilon$ it is expected that the transmission for frequency $k_n$ [recall Eq.~\eqref{eq7}] is not close to 1, meaning that the wave field $\psi_n$ is not at resonance.
Therefore, a natural step is to investigate the influence of $\epsilon$ on PTRs.

\section{Fate of PTRS under increasing perturbation strength}
\begin{figure}
\begin{center}
\includegraphics[width=1\columnwidth]{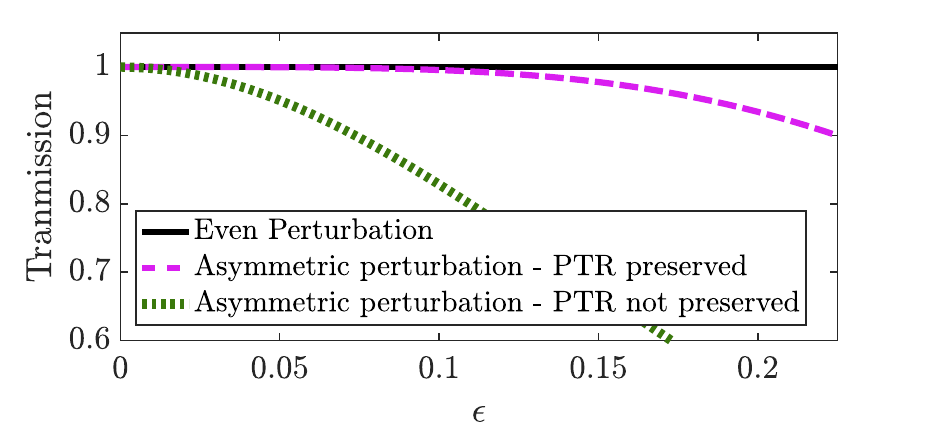}
\end{center}
\caption
{
The solid black line denotes the transmission of PTR number $n=1$ shown in Fig.~\ref{fig2}.
The dashed magenta (dotted green) line denotes the transmission of PTR number $n=1$ ($n=2$) shown in Fig.~\ref{fig4}(a).
}
\label{fig8}
\end{figure}

Before closing, we examine how the perturbation strength $\epsilon$ affects protected and unprotected PTRs by perturbations. First, we recall that to lose a PTR under mirror-\textit{symmetric} perturbations the eigenvalues of the $\mathcal{PT}$-symmetric problem given in Eq.~\eqref{eqboundary1} must coalesce (recall the discussion at the end of Sec. \ref{Section2}). We show in Fig.~\ref{fig8} with the solid line  such an example for the range of $\epsilon$ values for which the problem remains $\mathcal{PT}$ symmetric. Thus, the transmission remains exactly equal to 1 for these finite values of $\epsilon$.
In this symmetric case the perfect transmission will be lost at an exceptional point \cite{Davy2022}. Second, we show in Fig.~\ref{fig8} with a dashed line the transmission of a protected PTR by an \textit{asymmetric} perturbation. In this case, since $\text{Im}[k_1]=0$ in Eq.~(11), the effect of the perturbation at the PTR is of second order, and thus, the reflection coefficient increases as $r \sim \epsilon^2$. As a result, since the transmission coefficient is always given by $|t|^2=1-|r|^2$, we find that the dashed line in Fig.~\ref{fig8} drops as $1-|\epsilon|^4$. Finally, in Fig.~\ref{fig8} we also present the transmission of an unprotected PTR under the same asymmetric perturbation as in the latter case. Here, $\text{Im}[k_1] \neq 0$, and therefore, $r \sim \epsilon$, which in turn implies that the transmission of one of the unprotected PTRs drops as $1-|\epsilon|^2$.

\section{Concluding remarks}
\label{Section7}
In this paper, we explored how asymmetric perturbations affect the transmission efficiency of periodic systems composed of $N$ mirror-symmetric cells. 
To investigate this, we first applied classical perturbation analysis and derived an expression for minimizing reflection, which was found to have the form of an inner product between the wave function at resonance and the perturbed potential.
Then we introduced perturbations in the form of Dirac scatterers, aiming to preserve specific PTRs from the $N-1$ PTRs of the unperturbed system. 
Our analysis showed that using $q+1$ Dirac scatterers allows us to retain $q$ of the original PTRs. Additionally, we demonstrated that the mirror symmetry of the unit cell is crucial for preserving extra PTRs: When Dirac scatterers are placed at symmetry points within the unit cell, the preservation of one PTR necessarily implies the preservation of another.
We then examined the implications of the perturbation's asymmetry on the wave field, considering scattering from either the left or the right side of the perturbed setup. We demonstrated that the corresponding fields are not identical and exhibit asymmetry.
Finally, we showed that the impact of the perturbation strength on  unprotected PTRs is of order $\epsilon$, while its impact on protected PTRs is $O(\epsilon^2)$.

Our findings on achieving high transmission efficiency under asymmetric
perturbations open avenues for designing robust antireflection
coatings and reflectionless setups across multiple wavelengths. Extending these results to finite periodic quasi-1D systems or $\mathcal{PT}$-symmetric potentials with gains and losses could provide valuable insights into PTR robustness and nonreciprocal scattering \cite{Zhukovsky2010, Zhukovsky2011, McLeod1986, Biancalana2008, Taylor2011, Pagneux2017}. Exploring perturbing potentials without Dirac scatterers may also reveal novel phenomena, complementing the well-established study of wave scattering by point scatterers \cite{Holden1988, Lagendijk1998, Mostafazadeh2023}, which remains of both theoretical and experimental interest \cite{Kalozoumis2015}.

Moreover, our approach can be applied to matter waves in Bose-Einstein
condensates (BECs) governed by the Schrödinger equation in the
noninteracting regime. Noninteracting BECs can be achieved using Feshbach resonance to tune interatomic interactions \cite{Tiesinga2010}. Recent experimental advancements, such as guided matter waves and scattering in periodic
potentials \cite{Odelin2013, Bourdel2024}, align well with our model framework. With modern
techniques enabling the realization of arbitrary potential landscapes and subwavelength spatial features mimicking delta-function potentials \cite{Schmiedmayer2019, Rolston2018}, applying our treatment to ultracold atomic gases could inspire new experimental investigations.


\section*{Acknowledgements}
The authors acknowledge support from the Institute d'Acoustique - Graduate School of Le Mans.
I.K. acknowledges financial support from the Academy of Athens.
V.A. acknowledges financial support from the NoHeNA project funded under the program Etoiles Montantes of the Region Pays de la Loire. V.A. is supported by EU H2020 ERC StG “NASA” Grant Agreement No. 101077954.


\appendix

\section{Reflectionless scattering and $\mathcal{PT}$ symmetry}
\label{appendixA}

The eigenvalue problem of reflectionless modes can be written as
\begin{equation}
\dfrac{d^2\psi}{dx^2} + \left[ k^2 - \mathcal{V}(x) \right] \psi=0, 
\label{Appendix_Shrodinger}
\end{equation}
for $-D/2\leq x \leq D/2$, with the boundary conditions
\begin{equation}
\left. \dfrac{d\psi}{dx} \right|_{x=\pm D/2} = ik \psi (\pm D/2) . 
\label{Appendix_boundary1}
\end{equation}
Here,  $\psi$ is the eigenfunction, and $k$ is the eigenvalue which appears in the boundary condition. By manipulating Eqs.~\eqref{Appendix_Shrodinger} and \eqref{Appendix_boundary1} we find that, for a real and mirror-symmetric potential $\mathcal{V}$, if $\psi$ is an eigenfunction with eigenvalue $k$, then $\mathcal{PT}\psi$ is an eigenfunction with eigenvalue $k^*$, which is the signature of $\mathcal{PT}$ symmetry.

To make this $\mathcal{PT}$ symmetry even more explicit, let us note that, by using the auxiliary function $\chi \equiv k \psi$, we can write the eigenvalue problem of the reflectionless modes in the form
\begin{equation}
\hat{H}
\begin{pmatrix}
\psi  \\
\chi
\end{pmatrix}=
k \begin{pmatrix}
\psi  \\
\chi
\end{pmatrix},
\end{equation}
with the boundary conditions
\begin{equation}
\left. \dfrac{d\psi}{dx} \right|_{x=\pm D/2} -
i\chi(x=\pm D/2)=0,
\label{Appendix_boundary1A}
\end{equation}
where
\begin{equation}
\hat{H}=
\begin{pmatrix}
0 & 1   \\
-\dfrac{d^2}{dx^2} + \mathcal{V}(x) & 0
\end{pmatrix}.
\end{equation}

The advantage now is that the eigenvalue is no longer included in the boundary condition, and importantly, the latter is conserved by $\mathcal{PT}$ symmetry, $(\psi,\chi)^T$,  verifying Eq.~\eqref{Appendix_boundary1A} is equivalent to
$(\mathcal{PT}\psi,\mathcal{PT}\chi)^T$  verifying Eq.~\eqref{Appendix_boundary1A}.
Reflectionless modes are directly defined from the eigenvalues of the operator $\hat{H}$ acting on vector functions satisfying boundary condition \eqref{Appendix_boundary1A}. For the particular case of a real mirror-symmetric potential $\mathcal{V}$ we have 
\begin{equation}
\hat{H} \mathcal{PT} \begin{pmatrix}
\psi  \\
\chi
\end{pmatrix}=
\begin{pmatrix}
\chi^*(-x)  \\
-\dfrac{d^2 \psi^*}{dx^2}(-x) + \mathcal{V}(-x)
\end{pmatrix},
\end{equation}
with the boundary conditions $\psi'^*(\pm D/2)-i \chi^*(\pm D/2)=0$, and
\begin{equation}
\mathcal{PT} \hat{H} \begin{pmatrix}
\psi  \\
\chi
\end{pmatrix}=
\begin{pmatrix}
\chi^*(-x)  \\
-\dfrac{d^2 \psi^*}{dx^2}(-x) + \mathcal{V}(-x)
\end{pmatrix},
\end{equation}
with the boundary conditions $\psi'^*(\pm D/2)-i \chi^*(\pm D/2)=0$.

We can thus conclude explicitly, taking into account the boundary conditions, that
\begin{equation}
\hat{H} \mathcal{PT}=  \mathcal{PT} \hat{H},
\end{equation}
the commutation defining the $\mathcal{PT}$ symmetry of the eigenvalue problem. We can remark that the same $\mathcal{PT}$ symmetry of reflectionless modes would be obtained in the more general case of $\mathcal{PT}$-symmetric (complex) potential such that $\mathcal{PT}\mathcal{V}=\mathcal{V}\mathcal{PT}$.


\section{Perturbative expression}
\label{appendixB}

The wave function $\psi_{n}(x)$ and the frequency $k_{n}$ corresponding to the PTR number $n$ of the potential $\mathcal{V}(x)$
satisfy the Schr\"{o}dinger equation \eqref{Appendix_Shrodinger} and the boundary conditions \eqref{Appendix_boundary1}.

We write the potential $\mathcal{V}(x)$ in the perturbed form [Eq.~\eqref{perturbed_potential}]
\begin{equation}
\mathcal{V}(x)=\mathcal{V}_0(x)+\epsilon \mathcal{V}_1(x)
\label{perturbed_potential2}
\end{equation}
and the wave function $\psi_n$ and frequency $k_n$ as 
\begin{gather}
\psi_n=\psi_{0,n}+\epsilon \psi_{1,n}+\ldots  ,
\label{perturbed_field2}
\\
k_n=k_{0,n}+\epsilon k_{1,n} + \ldots ,
\label{perturbed_frequencies2}
\end{gather}
that is, Eqs.~\eqref{perturbed_field} and \eqref{eq7} respectively.
By plugging Eqs.~\eqref{perturbed_potential2}-\eqref{perturbed_frequencies2}
into Eqs.~\eqref{Appendix_Shrodinger} and \eqref{Appendix_boundary1} and collecting the terms in powers of $\epsilon$,
we find that $\psi_{0,n}$ and $k_{0,n}$ satisfy the zeroth-order problem [Eqs.~\eqref{Appendix_Shrodinger} and \eqref{Appendix_boundary1}].

After a few manipulations, we find that the correction $\psi_{1,n}$ satisfies the equation
\begin{small}
\begin{equation}
\dfrac{d^2 \psi_{1,n}(x)}{dx^2} +\left( k_{0,n}^2-\mathcal{V}_0 \right) \psi_{1,n} =- \left( 2k_{0,n} k_{1,n} -\mathcal{V}_1 \right) \psi_{0,n} 
\label{correction_field}
\end{equation}
\end{small}and the boundary conditions
\begin{footnotesize}
\begin{equation}
\left. \dfrac{d\psi_{1,n}(x)}{dx}\right|_{x=\pm D/2} - ik_{0,n} \psi_{1,n}(\pm D/2) = i k_{1,n} \psi_{0,n}(\pm D/2)~.
\end{equation}
\end{footnotesize}

By multiplying Eq.~\eqref{correction_field} by $\psi_{0,n}$,
integrating in the scattering region $[-D/2,D/2]$,
and using the equations that $\psi_{0,n}$ and $\psi_{1,n}$ satisfy, 
we arrive at  Eq.~\eqref{eqwave_number_corr} of the main text (the expression for $k_{1,n}$).


\section{Centers of the cells: Proof}
\label{AppendixC}
We give here the steps needed for the derivation of Eqs.~\eqref{eqReImcenters} and \eqref{eqgcenters}.
As we stated in the main text, these two equations hold for  the centers of the cells $a_p$, with $p=1,...,N$,
and for the case of PTRs (denoted throughout this work with the index $n$).

We begin by recalling that the cells were considered to possess mirror symmetry.
Due to the latter symmetry, the transfer matrix $\mathbf{M}$ of each unit cell [see  Eq.~\eqref{eqtransfer_one_cell}]
is decomposed around the axis of symmetry as $\mathbf{M}=\mathbf{M}_2 \mathbf{M}_1$, where 
\begin{equation}
\mathbf{M}_2^{-1}= \mathbf{P} \mathbf{M}_1 \mathbf{P}
\label{M2_1_2}
\end{equation}
and 
\begin{equation}
\mathbf{P}
=
\begin{pmatrix}
0 && 1 \\
1 && 0
\end{pmatrix}
\label{parity}
\end{equation}
is the parity operator.
Therefore, by setting \cite{Griffiths2001}
\begin{equation}
\mathbf{M}_1
=
\begin{pmatrix}
\alpha && \beta 
\\
\beta^* && \alpha^*
\end{pmatrix},
\label{M1}
\end{equation}
from Eqs.~\eqref{M2_1_2}-\eqref{M1} it follows that
\begin{equation}
\mathbf{M}_2^{-1}
=
\begin{pmatrix}
\alpha^* && \beta^*
\\
\beta && \alpha
\end{pmatrix}
.
\label{M2_1}
\end{equation}

As the next step, we define the vector $\boldsymbol{\chi}$ containing the right- and left-going waves at each point along the $x$ axis.
For scattering from the left side of the potential and for a frequency of the incident wave corresponding to the PTR $n$,
the following relation holds:
\begin{equation}
\boldsymbol{\chi}_n (-D/2)
=
\begin{pmatrix}
1 \\ 0
\end{pmatrix},
\label{vector_left}
\end{equation}
where $x=-D/2$ is the left edge of the scattering setup [see Fig.~\ref{fig1}(a)].
The wave function $\psi$ is the sum of the right- and left-going waves
(the sum of the two components of the vector $\boldsymbol{\chi}$),
and therefore, we get
\begin{equation}
\psi_n(-D/2)=1.
\label{point_D2}
\end{equation}

Next, we calculate the vector $\boldsymbol{\chi}_n$ at the right edge of the scattering setup, namely, at the point $x=D/2$.
To achieve this, we use the Chebyshev identity, which states 
that if the transfer matrix $\mathbf{M}$ is written in the form 
\begin{equation}
\mathbf{M}=
\begin{pmatrix}
m_{11} && m_{12} \\
m_{21} && m_{22}
\end{pmatrix},
\end{equation}
then the matrix $\mathbf{M}^N$ is given by
\begin{equation}
\begin{footnotesize}
\mathbf{M}^N=
\begin{pmatrix}
m_{11} U_{N-1}-U_{N-2} && m_{12} U_{N-1} \\
m_{12} U_{N-1} && m_{22} U_{N-1}-U_{N-2}
\end{pmatrix},
\end{footnotesize}
\end{equation}
where 
\begin{equation}
U_{N}=
\dfrac{\sin[(N+1)\phi]}{\sin(\phi)}
\end{equation}
and $\phi$ is the Bloch phase. 
For the case of a PTR, the Bloch phase is equal to $\phi_n=n\pi/N$ [recall Eq.~\eqref{eqBloch}], and therefore we get
\begin{equation}
\boldsymbol{\chi}_n(D/2)
=
\begin{pmatrix}
(-1)^n \\ 0
\end{pmatrix},
\end{equation}
meaning that 
\begin{equation}
\psi_n(D/2)=(-1)^n.
\label{pointD2}
\end{equation}

Subsequently, we calculate the wave function $\psi_n$ at point $a_1$ ($a_N$), namely, at the center of the first (last) cell.
Using Eqs.~\eqref{M1} and \eqref{vector_left}, we get 
\begin{equation}
\boldsymbol{\chi}_n(a_1)
=
\mathbf{M}_1 \boldsymbol{\chi}(-D/2)
=
\begin{pmatrix}
\alpha \\ \beta^*
\end{pmatrix}
,
\end{equation}
and therefore, the wave function $\psi_n$ at the center of the first cell is given by
\begin{equation}
\psi_n(a_1)=\alpha+\beta^*
.
\label{pointa1}
\end{equation}
Similarly, at point $a_N$
\begin{equation}
\boldsymbol{\chi}_n(a_N)
=
\mathbf{M}_2^{-1} \boldsymbol{\chi}_n(D/2)
=
\begin{pmatrix}
(-1)^n \alpha^* \\ (-1)^n \beta
\end{pmatrix},
\end{equation}
and as a result
\begin{equation}
\psi(a_N)=(-1)^n (\alpha^*+\beta)
.
\label{pointaN}
\end{equation}

We now proceed to calculate the wave function at the center of each cell $a_p$ (with $p=1,...,N$):
To accomplish this, first, we define the transfer matrix ${\mathbf{\tilde{M}}}=\mathbf{M}_1 \mathbf{M}_2$. 
Note that $\mathbf{\tilde{M}}$ differs from the transfer matrix $\mathbf{M}$ of a cell, which, as we noted before,
is equal to $\mathbf{M}=\mathbf{M}_2 \mathbf{M}_1$.
The two eigenvectors $\boldsymbol{v}_q$ and $\boldsymbol{v}_{-q}$ of ${\mathbf{\tilde{M}}}$ can be written in the following form [since $\det({\mathbf{\tilde{M}}})=1$]:
\begin{equation}
\boldsymbol{v}_q =
\dfrac{1}{\sqrt{1+|\kappa|^2}}
\begin{pmatrix}
1 \\ \kappa
\end{pmatrix}
,~~~
\boldsymbol{v}_{-q} =
\dfrac{1}{\sqrt{1+|\kappa|^2}}
\begin{pmatrix}
\kappa^* \\ 1
\end{pmatrix} 
,
\end{equation}
and $\kappa=\kappa^*$ due to the mirror symmetry of the cell; for a non-mirror-symmetric cell $\kappa \in \mathbb{C}$.
We now express the vector $\boldsymbol{\chi}_n(a_1)$ in the basis formed by $\boldsymbol{v}_q$ and $\boldsymbol{v}_{-q}$, 
\begin{equation}
\boldsymbol{\chi}_n(a_1) = A \boldsymbol{v}_q + B \boldsymbol{v}_{-q},
\label{pointa1_2}
\end{equation}
where the coefficients $A$ and $B$ are determined in the following step by the conditions given in Eqs.~\eqref{pointa1} and \eqref{pointaN};
the wave function at point $a_1$ is written in the form
\begin{equation}
\psi(a_1)=Az+Bz^*,
\end{equation}
where $z=\dfrac{1+\kappa}{\sqrt{1+|\kappa|^2}}$.

By acting on the vector $\boldsymbol{\chi}_n(a_1)$ given in Eq.~\eqref{pointa1_2} with the transfer matrix  ${\mathbf{\tilde{M}}}^{p-1}$,
we compute the vector $\boldsymbol{\chi}_n(a_p)$, and then we get the wave function at the center $a_p$ of a cell:
\begin{equation}
\boldsymbol{\chi}_n(a_p) = A e^{i(p-1)\phi_n} \boldsymbol{v}_q + B e^{-i(p-1)\phi_n} \boldsymbol{v}_{-q}
\end{equation}
and therefore,
\begin{equation}
\psi(a_p)=A e^{i(p-1)\phi_n} z + B e^{-i(p-1)\phi_n} z^*.
\label{psiam}
\end{equation}

The two coefficients $A$ and $B$ are found by using Eqs.~\eqref{pointa1} and \eqref{pointaN} and by setting $n=1$ and $n=N$ in Eq.~\eqref{psiam}.
The result is 
\begin{equation}
A=
\dfrac{w e^{-i(N-1)\phi_n}-w^* (-1)^n}{z \left( e^{-i(N-1)\phi_n} - e^{i(N-1)\phi_n} \right)} 
\label{coefficientA}
\end{equation}
and 
\begin{equation}
B=
\dfrac{(-1)^n w^* -w e^{i(N-1)\phi_n}}{z^* \left( e^{-i(N-1)\phi_n} - e^{i(N-1)\phi_n} \right)} ,
\label{coefficientB}
\end{equation}
where $w=\alpha+\beta^*$.
By substituting the expressions for $A$ and $B$  provided in Eq.~\eqref{coefficientA} and \eqref{coefficientB} into Eq.~\eqref{psiam},
we find that the wave function $\psi_n$ at the center $a_p$ of a cell is given by

\begin{equation}
\begin{split}
\psi (a_p) = & \frac{1}{\sin [(N - 1) \phi n]} \Big( w_R \sin [(N - p)\phi_n] \\
& + (-1)^n w_R \sin [(p - 1)\phi _n] 
 + i \big[ w_I \sin [(N - p)\phi _n] \\
& - (-1)^n w_I \sin [(p - 1)\phi _n] \big] \Big)
\label{wave_function_am}
\end{split}
\end{equation}
where $w_R=\text{Re}[w]$ and $w_I=\text{Im}[w]$.
After performing a few manipulations in  Eq.~\eqref{wave_function_am},
we find that 
\begin{equation}
\begin{split}
\text{Re}[\psi (a_p)] \text{Im}[\psi(a_p)] 
= &
\dfrac{w_R w_I }{\sin^2[(N-1)\phi_n]} 
\Big( 
\sin^2[(N-p)\phi_n] \\ & - 
\sin^2[(p-1)\phi_n]
\Big) .
\end{split}
\label{real_imag_center}
\end{equation} 
Equation~\eqref{real_imag_center} appears in the main text in the form of Eqs.~\eqref{eqReImcenters} and \eqref{eqgcenters}.


\section{Edges of the cells - Proof}
\label{AppendixD}
In order to derive Eqs.~\eqref{eqReImedges} and \eqref{eqgedges}, we follow an  approach similar to the one presented in Appendix \ref{AppendixC}
(we recall that the edges of the cells were denoted in the main text by $b_p$, $p=0,1,2,...,N$).

We note first that the two eigenvectors $\boldsymbol{u}_{\pm}$ of the transfer matrix $\mathbf{M}$ can be written in the following form:
\begin{equation}
\boldsymbol{u}_q =
\dfrac{1}{\sqrt{1+|\lambda|^2}}
\begin{pmatrix}
1 \\ \lambda
\end{pmatrix}
,~~~
\boldsymbol{u}_{-q} =
\dfrac{1}{\sqrt{1+|\lambda|^2}}
\begin{pmatrix}
\lambda^* \\ 1
\end{pmatrix} ,
\end{equation}
with $\lambda=\lambda^*$ due to the mirror symmetry of the cell.

We express the vector $\boldsymbol{\chi}_n(-D/2)=\boldsymbol{\chi}(b_0)$ in the basis formed by the vectors $\boldsymbol{u}_{\pm q}$,
\begin{equation}
\boldsymbol{\chi}_n(b_0)=
C 
\dfrac{1}{\sqrt{1+|\lambda|^2}}
\begin{pmatrix}
1 \\ \lambda
\end{pmatrix}
+D
\dfrac{1}{\sqrt{1+|\lambda|^2}}
\begin{pmatrix}
\lambda \\ 1
\end{pmatrix}
.
\label{pointb0}
\end{equation}
Since our analysis refers to PTRs, the vector $\boldsymbol{\chi}_n(b_0)$ is equal to
\begin{equation}
\boldsymbol{\chi}(-D/2)
=
\begin{pmatrix}
1 \\ 0
\end{pmatrix},
\label{pointb0_2}
\end{equation}
and by equating Eq.\eqref{pointb0} with Eq.\eqref{pointb0_2} we calculate the coefficients $C$ and $D$.
We find that
\begin{equation}
\boldsymbol{\chi}(b_0)=
\dfrac{1}{1-\lambda^2}
\begin{pmatrix}
1 \\ \lambda
\end{pmatrix}
-
\dfrac{1}{1-\lambda^2}
\begin{pmatrix}
\lambda \\ 1
\end{pmatrix}
.
\label{pointb0n}
\end{equation}

As a next step, we compute the vector $\boldsymbol{\chi}_n(b_p)$ by acting with the transfer matrix $\mathbf{M}^p$ on the vector $\boldsymbol{\chi}_n(b_0)$:
\begin{equation}
\boldsymbol{\chi}_n(b_p)=
e^{ip\phi_n}
\dfrac{1}{1-\lambda^2}
\begin{pmatrix}
1 \\ \lambda
\end{pmatrix}
-
e^{-ip\phi_n}
\dfrac{1}{1-\lambda^2}
\begin{pmatrix}
\lambda \\ 1
\end{pmatrix},
\end{equation}
where, again, $\phi_n=n\pi/N$ is the Bloch phase of the $n$th PTR.

The sum of the two components of the vector $\boldsymbol{\chi}_n(b_p)$ is the wave function $\psi_n(b_p)$, namely
\begin{equation}
\psi_n(b_p)
=
\dfrac{1}{1-\lambda} \left[ e^{ip\phi_n} -\lambda e^{-ip\phi_n}\right]
.
\label{psi_edges}
\end{equation}
Finally, from Eq.~\eqref{psi_edges} we find that the quantity $\text{Re}[\psi_n(b_p)] \text{Im}[\psi_n(b_p)]$ is given by
\begin{equation}
\text{Re}[\psi_n(b_p)] \text{Im}[\psi_n(b_p)]
=
\left(
\dfrac{1}{2}
\right)
\left(
\dfrac{1+\lambda}{1-\lambda}
\right)
\sin(2p \phi_n),
\end{equation}
corresponding to  Eqs.~\eqref{eqReImedges} and \eqref{eqgedges}.




\begin{thebibliography}{99}
\bibitem{comment1} We do not display the case where the PTRs $n=2$ and $N-n=6$ are simultaneously preserved because the quantity $\text{Re}[\psi] \text{Im}[\psi]$ is zero at the edge $b_2$ for the corresponding frequency, suggesting that the strength $c_1$ of the first Dirac scatterer has to be zero in order to preserve the PTR $n=2$ (or $n=6$) [recall Eq.~\eqref{eq_wave_number__Dirac_preserved}]. However, Fig.~\ref{fig5}(a) does not represent such a scenario; namely, only one Dirac scatterer is placed  at the unperturbed setup. We have verified, however, that if we add only one Dirac scatterer at edge $b_2$, then the PTRs $n=2,6$ are simultaneously preserved.
\end{thebibliography}
\end{document}